\magnification=\magstep1 

\font\sf=cmss10
\hsize=6truein
\def \d{{\partial}}
\def \arcch {{\rm arccosh}}
\def \um{{\textstyle {1\over2}}}
\def \tm{{\textstyle {3\over2}}}
\def \uq{{\textstyle {1\over4}}}
\def \i{{\rm i}}
\def \e{{\rm e}}
\def \f{{\rm f}}

\tenrm 
\hfuzz 15pt

\centerline{ GRAVITON EMISSION AND LOSS OF COHERENCE }

\vskip 2pc
 \centerline {Giorgio Calucci \footnote *{E-mail:{\tt giorgio@ts.infn.it}} }
 \vskip 1
 pc
\centerline{{\it Dipartimento di Fisica teorica dell'Universit\`a di Trieste,
 I 34014}}

\centerline{{\it INFN, Sezione di Trieste, Italy}}
\vskip 3pc

\centerline{\bf {Abstract} } 
\vskip 1pc
{\narrower\narrower{
The decoherence effect due to emission of gravitons is examined. It shows the 
same qualitative features of the QED effect which has already been investigated,
it is obviously much weaker, wholly universal and shows a stronger energy 
dependence. The result can be extended to photons, they also may undergo 
decoherence due to graviton emission.
For this limited aim the incomplete status of the quantum gravity, 
in comparison with QED, is not source of severe difficulties because all the
effects are attributed to the infrared sector of the dynamics.}}

\vfill
\eject
{\bf I. Introduction}
\vskip 1pc  
The loss of quantum coherence due to bremsstrahlung process has been 
studied in detail [1,2] for the electromagnetic case. The electromagnetic
radiation is certainly present whenever charged particles undergo scattering
processes, so, without assuming that the radiation is the unique process giving
rise to the loss of quantum coherence, we may be sure that electrodynamical
effects are there. It is also evident that this kind of decoherence mechanism 
makes a sharp distinction between charged and uncharged particles, it becomes 
therefore natural to look for other kinds of radiation that may implement the 
same effect in a universal way. 
The obvious candidate is the gravitational radiation since the
gravity is the most universal form of interaction known at present, so in this 
note the effects of gravitational bremsstrahlung are examined.
Some possible source of difficulties, appears 
immediately: when we investigate the effects of photon emission we have at our 
disposal QED
which probably the best settled (and the oldest) branch of quantum field theory,
the quantum theory of gravitation still waits a final shape. The small sector of
the general quantum mechanics of gravitation that concerns the infrared
radiation can be treated in close analogy with QED as it can be seen from the
general treatment of the infrared radiations  given by Weinberg [3]. This
treatment will be systematically followed in this note, some points will be 
recalled or re-elaborated when necessary. The difference that cannot be
neglected is the fact that the emission of photon is due to the electric charge
which is a Lorentz scalar and that it is quite possible to consider situations
where every interacting particle keeps its charge, the emission of gravitons is
due to the energy-momentum four vector and that this quantity is certainly not
kept by the single interacting particle although it is conserved in the overall
process. This fact is conceptually obvious but gives rise to some technical
difficulties: for this reason in this paper the perturbative treatment is
presented formerly, in sect. II because there the effects of the overall 
conservation are
better shown, after an extension of the Bloch-Nordsieck model in proposed in
sect.III where
one sees that the effect of decoherence can be studied following each
particle individually.
The fact that also photons may undergo decoherence due to gravitational 
bremsstrahlung is noted.
\vskip 1pc
{\bf II. Infrared gravitons}
\vskip 1pc
{\bf A. Infrared compensation for standard states}
\vskip 1pc
Some results given in ref.[3] will be explicitly needed, so they are now
recalled and adapted to the particular problem under investigation. Since the 
source of the gravitons is the energy-momentum
tensor one must consider the whole scattering process, where this source is
conserved, it is not enough to follow a particular line as it is allowed in
QED [5]. To avoid unessential complications the scattering process is an elastic
two-body collision, the masses are taken to be equal.\par
The units $c=\hbar=1$ and the coupling $\kappa=\sqrt {8\pi G}$ are used ($G$ is
the Newtonian constant), the Minkowski metric is denoted by $\eta_{\mu\nu}$, 
the Mandelstam invariant $s,t,u$ are also employed. \par
If one calls ${\cal M}_o$ the amplitude for the scattering without further
radiation \par
{\sf Fig.1}\par
 then one obtains the amplitude for the scattering with the radiation of a
soft graviton [3]: \par {\sf Fig.2} 
 $${\cal M}_r={\cal M}_o\cdot \i\kappa 
 \Big[{{q^{\mu}q^{\nu}}\over {q\cdot k}}+{{q'^{\mu}q'^{\nu}}\over {q'\cdot k}}-
 {{p^{\mu}p^{\nu}}\over {p\cdot k}}-{{p'^{\mu}p'^{\nu}}\over {p'\cdot k}}\Big]
 (2\pi)^{-3/2} f_{\mu\nu} \eqno(1)$$
 One can get this expression working out $e.g.$ the soft limit of the emission
 by a spinorial particle, using the vertex[6]\footnote*{In that paper the
 $\kappa$ is twice the $\kappa$ used here.}:
 $$\um \i \kappa\{\um [\gamma^{\mu}(p_1+p_2)^{\nu}+\gamma^{\nu}(p_1+p_2)^{\mu}]+
    \eta^{\mu\nu}[(\hat p_1+\hat p_2)-2m]\}$$
  but it is quite general, it does not depend on the spin of the emitting
  particle. The polarization vector is $f_{\mu\nu}$ and since the source is
  conserved the sum over polarization is simply given by: 
  $$\sum_i f^i_{\mu\nu}f^i_{\rho\sigma}= \Pi_{\mu\nu\rho\sigma}=\um 
 (\eta_{\mu\rho}\eta_{\nu\sigma}+\eta_{\nu\rho}\eta_{\mu\sigma}-
 \eta_{\mu\nu}\eta_{\rho\sigma}) \eqno(2)$$
 So the transition probability, with emission of an infrared graviton is given 
 by

 $$\eqalign{{\cal X}=
 {1\over {(2\pi)^3}}|{\cal M}_o|^2&
 \int {{d^3k}\over {2\omega}}\kappa^2\bigg[\um m^4
 \Big({1\over{(q\cdot k)^2}}+{1\over{(q'\cdot k)^2}}+
      {1\over{(p\cdot k)^2}}+{1\over{(p'\cdot k)^2}}\Big)+\cr
      &\Big({{2(q\cdot q')^2-m^4}\over {q\cdot k q'\cdot k}}+
            {{2(p\cdot p')^2-m^4}\over {p\cdot k p'\cdot k}}-
            {{2(q\cdot p)^2-m^4}\over {q\cdot k p\cdot k}}-\cr
           &{{2(q\cdot p')^2-m^4}\over {q\cdot k p'\cdot k}}-
            {{2(p\cdot q')^2-m^4}\over {p\cdot k q'\cdot k}}-
            {{2(p'\cdot q')^2-m^4}\over {p'\cdot k q'\cdot k}}\Big)\bigg]}
            \eqno(3)$$
  The notation ${\cal X}_1$ will be used to indicate the sum of the first four
  terms, while ${\cal X}_2$ will be used to indicate the sum of the last six
  terms. 
  The integration over the energy of the radiated graviton goes from a minimum
  $\lambda$, and a maximum $\Lambda$ that has no role in the discussion, but for
  allowing the low-energy approximation. Performing the integration the
  following result is obtained: 
  $${1\over {2\pi^2}}|{\cal M}_o|^2
    \kappa^2 m^2
  \big[1+D(p\cdot p'/m^2)-D(p\cdot q/m^2)-D(p'\cdot q/m^2)\big] 
  \ln (\Lambda/\lambda)\eqno (4)$$
  where
  $$D(x)={{2x^2-1}\over\sqrt{x^2-1}}\arcch\,x \qquad D(1)=1\qquad 
  D(x)\to 2x\ln 2x\; {\rm for} \; x\to \infty. \eqno(5)$$ 
  Now one must look at the virtual corrections: 
  there are six terms corresponding to the correction in the channels $s,t,u$
  respectively,\par {\sf Fig.3}\par
  working always in the i.r. limit the corresponding amplitude
  may be expressed as:
   $${\cal M}_v=2[{\cal M}_s+{\cal M}_t+{\cal M}_u]\eqno (6)$$
 and the correction, at the order $\kappa^2$ is given by 
 $2\Re({\cal M}_o{\cal M}^*_v)$
   $${\cal M}_s={{-\i}\over {(2\pi)^4}}\kappa^2{\cal M}_o
 \int {{d^4 l}\over {l^2+\i\epsilon}}\;{{(p\cdot p')^2-\um m^4}\over 
 {(p\cdot l-\i\epsilon)(p'\cdot l+\i\epsilon)}}\eqno(7a)$$
  $${\cal M}_t={{\i}\over {(2\pi)^4}}\kappa^2{\cal M}_o
 \int {{d^4 l}\over {l^2+\i\epsilon}}\;{{(p\cdot q)^2-\um m^4}\over 
 {(p\cdot l-\i\epsilon)(q\cdot l-\i\epsilon)}}\eqno(7b)$$
 $${\cal M}_u={{\i}\over {(2\pi)^4}}\kappa^2{\cal M}_o
 \int {{d^4 l}\over {l^2+\i\epsilon}}\;{{(p\cdot q')^2-\um m^4}\over 
 {(p\cdot l-\i\epsilon)(q'\cdot l-\i\epsilon)}}\eqno(7c)$$
 The integration over the virtual momenta which takes into account the 
 contribution of the photon
 pole gives the result needed to cancel one part of the real i.r.
 divergence [4,5], the term ${\cal X}_2$ .
 The contribution of the second pole of the s-channel term is easily calculated
 in the frame ${\vec p'}=0$, where it corresponds to $l_o=-\i\epsilon$ with the 
 result
 $${\cal M}^{(2)}_s=={\i\over {4\pi}}\kappa^2{\cal M}_o
[(p\cdot p')^2-\um m^4]{1\over {m|{\vec p}|}}\int {{d\ell}\over {\ell}}\qquad 
\ell=|{\vec l}| \eqno(7d)$$
 This term is imaginary, so it does not contribute to the correction to order
 $\kappa^2$, it corresponds (in QED) to the perturbative expansion 
 of the Coulomb phase[4,7],  here it correspond to the gravitational elastic
 rescattering of the incoming or outgoing particles. \par
 There are still some other contributions, that are needed to cancel the 
 divergent terms in ${\cal X}_1$. Here a difference with respect to QED is found
 , in fact in QED either
 one chooses the Yennie gauge [8], so that the virtual corrections till now
 considered will completely cancel the i.r. divergent part coming from real
 radiation, or, when using the Lorentz gauge, one look at the
 i.r. divergent part of the renormalization constant $Z_2$ for the wave function of
 the charged particle. Here there is no simple analogous of the Yennie gauge and
 a general renormalization procedure is not available. However when a self
 energy correction is inserted with the procedure:
 $${1\over {\hat p-m}}\to {1\over {\hat p-m}}\Sigma (p) {1\over {\hat p-m}}$$
  one must keep in the limit $p\to \bar p\quad \bar p^2=m^2$, i.e. when the
  insertion is performed on one external leg, the correct position of the mass
  pole and the correct residuum at the pole,\par 
 {\sf Fig.4}\par
this amount to two subtractions that, in the i.r. regime are well defined; in
 particular the second subtraction term contains the factor
 $\d\Sigma/\d p \big|_{p\to\bar p}$, when the derivative acts on the numerator 
 of $\Sigma$ it does not yield i.r. divergences, because $\Sigma$ itself is not 
 i.r. divergent, on the denominator the derivative acts in the same way 
 as in QED, and yields therefore the divergent terms needed to cancel the term 
 ${\cal X}_1$. The important, and unfortunate, difference is that now these and 
 similar subtractions does not avoid the ultraviolet divergences, but for our
 limited aim only the infrared sector matters.
  \par
  In simple formulation the result is that the i.r. radiation has a matrix
  element
  $${\cal M}_r(k)={\cal M}_o\cdot \kappa\beta(k)\eqno (8)$$
  The virtual correction has an i.r. contribution 
 $${\cal M}_v={\cal M}_o\cdot\kappa^2\rho_v\eqno(9)$$
 and that
 $$\int |\beta(k)|^2 {{d^3k}\over {2\omega}}+2\Re\rho_v=C\eqno(10)$$
 The term $C$ is finite, in the true i.r. limit is zero.
  \vskip 1pc
  {\bf B. Superposition of two states of motion}  
\vskip 1pc
  If the two-particle final state is the superposition of two states of motion, like
  $|{\cal F}>=[|q_1,q'_1>+|q_2,q'_2>]/\sqrt {2}$, then the transition
  probability without any soft particle correction is:
  $$\um |{\cal M}_o^{(1)}+{\cal M}_o^{(2)}|^2\eqno (11)$$
  The transition probability with emission of one soft particle of momentum $k$
  is:
  $$\um |{\cal M}_o^{(1)}\cdot \kappa\beta^{(1)}(k)
        +{\cal M}_o^{(2)}\cdot \kappa\beta^{(2)}(k)|^2\eqno (12a)$$
  The transition probability including the correction for one virtual
  soft particle is: 
   $$\um |{\cal M}_o^{(1)}\cdot(1+\kappa^2\rho_v^{(1)})
         +{\cal M}_o^{(2)}\cdot(1+\kappa^2\rho_v^{(2)})|^2\eqno (13a)$$ 
  so the virtual correction to the order $\kappa^2$ may be written as:
      $$C_v=\kappa^2\big[ 
  |{\cal M}_o^{(1)}|^2\rho_v^{(1)})+|{\cal M}_o^{(2)}|^2\rho_v^{(2)})+
      2\Re({\cal M}_o^{(1)}{{\cal M}_o^{(2)}}^*)
      (\rho_v^{(1)}+\rho_v^{(2)})\eqno (13b)$$
      while the total infrared real correction is:  
   $$C_r=\um \int\big[|{\cal M}_o^{(1)}\beta^{(1)}(k)|^2
                 +|{\cal M}_o^{(2)}\beta^{(2)}(k)|^2+
   2\Re({\cal M}_o^{(1)}\beta^{(1)}(k){{\cal M}_o^{(2)}}^*\beta^{(2)}(k))\big]
       {{d^3k}\over {2\omega}}\eqno (12b)$$
       
  By using the compensation discussed previously between real and virtual
  correction we find that the only a surviving addendum contains the 
  interference term.  
  
 $$C=C_r+C_v=-\kappa^2 \Re({\cal M}^{(1)}_o {\cal M}^{*(2)}_o)\int 
 [\beta^{(1)}(k)-\beta^{(2)}(k)]^2 {{d^3k}\over{2\omega}}\eqno (14)$$
 This expression can be recast in a form similar to eq.(4)
  $$\eqalign{
  C=&-2\pi\kappa^2 m^2  \Re({\cal M}^{(1)}_o {\cal M}^{*(2)}_o)\cr 
  &\big[1+D(q_1\cdot q'_1/m^2)-D(q_1\cdot q_2/m^2)-D(q_1\cdot q'_2/m^2)\big] 
 \ln (\Lambda/\lambda)\;.} \eqno (15)$$
 If the two final states $|q_1,q'_1>$ and $|q_2,q'_2>$ which build up the
 superposition are not very different, it is convenient to use the expansion:
  $$q_2=q_1+\delta q \;\quad q'_2=q'_1-\delta q$$   
  In the centre-of-mass frame the common three momentum is $Q^2=\uq s-m^2$ and
   $\delta q\cdot q_1=-\delta q\cdot q'_1=2Q^2\sin^2\phi/2$, being $\phi$ the
   angle, that is now assumed small, between the two final directions $q_2$ and 
   $q_1$
    $$C=-4\pi\kappa^2 \Re({\cal M}^{(1)}_o {\cal M}^{*(2)}_o) Q^2\sin^2\phi/2 
 \big[\dot D(p\cdot p'/m^2)-\dot D(1)\big]\ln (\Lambda/\lambda)\;.\eqno (16a) $$ 
 From previous expressions
  $$\dot D(1)=\tm\qquad 
  \dot D(x)\to 2 (1+\ln 2x)\quad {\rm for}\quad x\to\infty \eqno (16b)$$  
  When the radiating particle is massless some minor modification are required:
  if $m\to 0$ then $x\to\infty$ in eq. (5) so the asymptotic form of $D(x)$ is
  certainly correct 
  $$\eqalign{
  C= -2\pi\kappa^2 &\Re({\cal M}^{(1)}_o {\cal M}^{*(2)}_o) 
  \big[m^2+2q_1\cdot q'_1 \ln (2q_1\cdot q'_1/m^2)-\cr
           &2q_1\cdot q_2 \ln (2q_1\cdot q_2/m^2)-
           2q_1\cdot q'_2 \ln (2q_1\cdot q'_2/m^2)\big]
   \ln (\Lambda/\lambda) }$$
  or also
  $$\eqalign{
  C= &-4\pi\kappa^2\Re({\cal M}^{(1)}_o {\cal M}^{*(2)}_o) \Big( 
     \big[q_1\cdot q'_1 \ln (2q_1\cdot q'_1/s)-
           q_1\cdot q_2 \ln (2q_1\cdot q_2/s)-\cr
         & q_1\cdot q'_2 \ln (2q_1\cdot q'_2/s)\big]+
    \big[(q_1\cdot q'_1 -q_1\cdot q_2 -q_1\cdot q'_2 ) 
           \ln (s/m^2)\big]\Big) \ln (\Lambda/\lambda) }\eqno (17) $$
  Energy-momentum conservation gives $q_1+q'_1-q_2-q'_2=0$ so that in the limit
 $q_1^2=m^2\to 0$ the second parenthesis goes to zero.  \par
 In analogy with eq. (16) it is possible to give the expression for small $\phi$
    $$C= 4\pi\kappa^2\Re({\cal M}^{(1)}_o {\cal M}^{*(2)}_o) 
      2Q^2 \sin^2\phi/2[2\ln\sin\phi/2-1]\eqno (18) $$
  Comparing this expression with eq.(16a) it appears that the dependence on
  the presence of a logarithmic term is the only remnant of the massless 
  condition, at least for small angles.
   The fact the it is possible to deal with the i.r. divergence coming from
   massless particles emitted by a massless source in a simple way is, 
   as shown in [3], a peculiarity of the graviton emission, with
   photon emission the situation would be much worse; experimentally we know
   that there are massless particle having gravitational interactions, the
   photons, but there are not massless charged particles and if we look for
   charged massless particles in quantum chromodynamics we find indeed a
   very complicated infrared behaviour.   
 \vskip 1pc   
{\bf  III. Explicit description of the time evolution}
\vskip 1pc  
{\bf A. The Bloch and Nordsieck model reviewed}
\vskip 1pc 
The usual tools of scattering theory with emission of soft massless bosons
going back to Bloch and Nordsieck model will be used[4], in that form which has
been reviewed in the previous treatment of the QED effects[2], so only the 
points that show significant differences will be presented. The main difference
is that, as discussed in section II, it is not obvious that it is enough
to follow only one particle in its fly, if we look to the final state, one 
should take into account both outcoming particles and together with their 
radiation and see how the compensation between real and virtual corrections
is realized in this case. The following analysis shows, however, that the
compensation happens in transparent way so that the consideration of only one
particle, together with its radiation is, at the end, justified.
The scattering process (for particles of equal masses) is described
by the Hamiltonian [3]: 
$$ H=H_o+V=H_o+\sum_{J=1}^2V_J+\Delta \eqno(19a)$$
$$ H_o= -\i\vec v_1\cdot \vec \partial_1 + {m\over{\gamma_1}}
        -\i\vec v_2\cdot \vec \partial_2 + {m\over{\gamma_2}} + 
  \sum_{\nu} \int {d^3 k}\omega a_{k,\nu}^{\dagger}a_{k,\nu}\eqno(19b)$$
 $$V_J=-{\kappa m\over {(2\pi)^{3/2}}}
         \sum_{\nu} \int {{d^3 k}\over {\sqrt{2 \omega}}}\gamma_J
            \bigl[a_{k,\nu} \vec v_J \cdot \f_{k,\nu}\cdot \vec v_J
            \e^{\i \vec k \cdot \vec r_J}+
        a_{k,\nu}^{\dagger} \vec v_J \cdot  \f_{k,\nu} \cdot\vec v_J 
     \e^{-\i \vec k \cdot \vec r_J} \bigr]\;.\eqno(19c)$$
  The polarization tensors for the gravitons are expressed here in non-covariant 
  Coulomb gauge and satisfy the well-known relations
  $$\sum_{\nu} \f_{k,\nu}^{cd}\f_{k,\nu}^{ab}=
  \um[u_{ac}u_{bd}+u_{ad}u_{bc}-
    u_{ab}u_{cd}]\,\quad u_{ij}=\delta_{ij}-\hat k_i \hat k_j.$$ 
    The Latin indices
    are purely spatial and $\hat k$ is the unit vector $\vec k/\omega$.
     The constant $\Delta$ is an energy renormalization term, having the role
    that in a covariant treatment is played by mass counter-term. \par
 It has been already shown[2] that the evolution operator takes the form   
       $${\cal U}(t)=
\exp\bigg[-\um\int_o^t\!\int_o^t D(\tau,\tau')d\tau,d\tau'\bigg]
{\cal N}\exp \int_o^t [-\i \tilde V(\tau) d\tau]\,,\eqno(20)$$
with $\tilde V(t)=\e^{\i H_ot}V \e^{-\i H_ot}$. \par
 Due to the actual form of $\tilde V(t)$, in the present case we have  
 $$D(\tau, \tau')=\sum_{I,J}D_{I,J}(\tau, \tau')\eqno(21a)$$
 and   
$$\eqalign{
D_{I,J}(\tau, \tau')=&<|{\cal P}(\tilde V_I(\tau)\tilde V_J(\tau'))|>
={{\kappa^2 m^2}\over {(2\pi)^3}}
\sum_{\nu}\int {{d^3 k}\over {2\omega}}
\gamma_I\gamma_J\vec v_I \cdot  \f_{k,\nu} \cdot\vec v_I
                \vec v_J \cdot  \f_{k,\nu} \cdot\vec v_J\cr
&[\e^{\i (\vec k \cdot (\vec r_I-\vec r_J)} 
 \e^{\i (\omega-\vec k \cdot \vec v_J)\tau'}
 \e^{-\i (\omega-\vec k \cdot \vec v_I)\tau}
 \vartheta(\tau-\tau')+(\tau,I) \leftrightarrow (\tau',J)]\,}\eqno(21b)$$
From the form of $\cal U$ one sees that the radiation of one soft particle
involves the absolute square of the matrix element 
$<|\tilde V_I(\tau)|\vec k,\nu>$:
which is made up by addenda of the form:
    $$\eqalign{&
A_{I,J}(\tau, \tau')=
<|\tilde V_I(\tau)|\vec k,\nu><\vec k,\nu|\tilde V_J(\tau')|>\cr
&={{\kappa^2 m^2}\over {(2\pi)^3}}
\sum_{\nu}\int {{d^3 k}\over {2\omega}}
\gamma_I\gamma_J\vec v_I \cdot  \f_{k,\nu} \cdot\vec v_I
                \vec v_J \cdot  \f_{k,\nu} \cdot\vec v_J
[\e^{\i (\vec k \cdot (\vec r_I-\vec r_J)} 
 \e^{\i (\omega-\vec k \cdot \vec v_J)\tau'}
 \e^{-\i (\omega-\vec k \cdot \vec v_I)\tau}]}\eqno(22)$$
 It is only a matter of calculation to verify that 
 $\Re \int_o^t\int_o^t d\tau d\tau'D=\int_o^t\int_o^t d\tau d\tau'A$. \par
 For the terms
 where $I=J$ this is precisely the result already used in [2], the exponential
  of
 $-\Re \int_o^t\int_o^t d\tau d\tau'D$ coming from eq.(20) is precisely
 compensated by the exponentiation of $\int_o^t\int_o^t d\tau d\tau'A$, which
 comes from the sum over the radiated bosons, when they build up a coherent
 state, the only difference comes from the polarization factors. The case
 $A_{1,2}+A_{2,1}$ involves explicitly the 
 difference $\vec r_2-\vec r_1$ through the term 
 $R=\hat k\cdot (\vec r_2- \vec r_1)=\omega(r_2 z_2-r_1 z_1)$, having introduced
the cosines, $z_J=\cos \theta_J$, of the angles between $\vec k$ and $\vec v_J$.
\par
 The actual expression we get for  ${\cal A}=\int_o^t\int_o^t d\tau
 d\tau'[A_{1,2}+A_{2,1}]$ is:
 $$ \eqalign{&{\cal A}={{\kappa^2 m^2}\over {(2\pi)^3}}\int d\Omega\um d\omega
     {{v_1^2 v_2^2\gamma_1\gamma_2 (1-z_1^2)(1-z_2^2)}
     \over {(1-v_1z_1)(1-v_2z_2)}}\times\cr
     &\Big[ \cos \omega R
     -\cos \omega[R-(1-v_1z_1)t]-\cos \omega[R+(1-v_2z_2)t]
     +\cos \omega [R+v_1z_1t-v_2z_2t]\Big]} \eqno(23)$$
     and coincides with $\Re\int_o^t\int_o^t d\tau
 d\tau'[D_{1,2}+D_{2,1}]$. So here also there is a compensation between virtual
 and real contributions.\par
 The term coming from $\Im D$, which 
 remains there and cannot be eliminated by some renormalization procedure just
 because it
 depends on $\vec r_2-\vec r_1$, is a part of the rescattering phase: the phase 
 that in QED, we would call Coulomb phase 
 \footnote*{It is not, however the 
  complete rescattering phase: since the treatment is not covariant there is a
  contribution coming from the static gravitational interaction; in the
  covariant perturbative treatment, shown in sect. II, these two terms are
  never separated.}.\par
   We are in any case interested in the behaviour for large $t$, so there is a
  question on how we deal with $R$. The simpler attitude, followed in [2], was
  to think of $\vec r_J$ as independent of $t$, plane wave for the outgoing
  particles, if however we think, more realistically, that we have wave packets,
  although very broad, we would take, say, $\vec r_J= \vec s_J+\vec v_J t$, with
  fixed $ \vec s_J$ in the wave functions of the scattered particles, and so 
  $R=S+(v_2z_2-v_1z_1)t$, we see, however, that this substitution would amount
  in the previous expression eq.(23) to a substitution of $\vec v_J$ with 
  $-\vec v_J$ and of 
  $R$ with $S$, this will not affect the $t\to\infty$ limit.\par
  These calculations show what really one expected i.e. that the
  Bloch-Nordsieck cancellation works also for two interacting outgoing
  particles;
  more precisely, that the diagonal terms compensates among themselves and the
  same do the cross terms.
  So now the study of the evolution of the superposition of two states of motion
  can be done, as
  in QED, looking only at one outgoing particle.

\vskip 1pc    
 {\bf B. Superposition of two states of motion} 
 \vskip 1pc   
 
  There are two differences with the QED: one is that instead of $e$ we
  find $\kappa m\gamma$, the second lies in the polarization term, but once we
  look at the evolution of one final particle with its radiation all the
  qualitative features remain the same as in QED, so only the initial
  definitions and the differences with
  respect to the previous case will be discussed. 
  If the state after the scattering is a superposition of two states
 characterized by velocities $v_=$ and $v_-$ there are two transition amplitudes
  generated by the evolution operator ${\cal U}(t)$, say ${\cal T}_+$ and 
  ${\cal T}_-$
  and then an interference term is produced in the transition probability at 
  finite time:
  $$ {\cal I}(t)=\Re\big[{\cal T}_-^*{\cal T}_+\big]$$
  Each of the two factors carries its virtual correction term as discussed
  before and in [2], but the real emission of soft bosons gives, for each boson
  which is emitted, a factor:
  \footnote*{the real part of eq.(24) has the same form as the integrand of eq.
  (23), the quantities entering in them have however different meaning: here 
  $v_+$ and
  $v_-$ refer to two different states of motion of the same particle, in eq.(23)
  $v_1$ and $v_2$ refer to the two final particles.}:
   $$\eqalign{
 q(\omega_R)&={{(\kappa m\gamma)^2}\over{(2\pi)^3}}\int d\Omega 
 {{\sum_{pol}(\vec v_+ \cdot \f\cdot\vec v_+)
 (\vec v_- \cdot \f\cdot\vec v_-)}\over
 {(\vec k\cdot \vec v_+-\omega)(\vec k\cdot \vec v_- -\omega)}}\omega d\omega\cr
 &\big[1-\exp[-\i(\vec k\cdot \vec v_+-\omega)t]-
 \exp[\i(\vec k\cdot \vec v_- -\omega)t]+
 \exp[\i(\vec k\cdot \vec v_- -\vec k\cdot \vec v_+)t]\big]\;. }\eqno(24)$$
 Summing over all the emitted bosons this term is exponentiated, whereas the
 virtual correction terms are noting but the negative exponentials of $\um q$,
  once with all the velocities equal to $\vec v_+$
  and once with all the velocities equal to $\vec v_-$.\par
 The long time evolution produces also here a dominant term in $\ln(\omega_R t)$
 whose coefficient $X=\int d\Omega \xi$ can be calculated starting from eq.(24);
 assuming here also that the two speeds are equal and calling $\delta$ the angle
 between the two directions it results:
 $$\xi={{G m^2\gamma^2}\over {\pi^2}}v^4
 {{(\cos
 \delta-\cos\theta_+\cos\theta_-)^2-\um(\sin\theta_+\sin\theta_-)^2}\over
 {(1-v\cos\theta_+)(1-v\cos\theta_-)}} \eqno(25)$$
 This expression could have been directly obtained from the corresponding one
 for
 the QED case with the appropriate substitutions of the coupling constant and 
 of the spin projector, all the previous part of Sect. III may be interpreted as
 a justification of eq.(25).\par
 At this point we must extract some simpler information out of the general form.
 \par
 In the limit $\delta\to 0$ one gets 
 $$\xi_o={{G m^2\gamma^2}\over {2\pi^2}}v^4 {{(\sin\theta)^4}\over
 {(1-v\cos\theta)^2}}$$
 so that 
 $$ X_o ={{G m^2\gamma^2}\over {\pi^2}} {2\over v}
 \Big[2v-{4\over 3}v^3-(1-v^2)\ln {{1+v}\over{1-v}}\Big]\;. \eqno(26)$$
 This expression has, in turn, the two limits
 $$ v\to 1 \quad  X_o  \to {{8G m^2\gamma^2}\over {3\pi}}\qquad v\to 0
 \quad X_o  \to {{G m^2}\over {\pi}}v^4{{16}\over{15}}\Big[1+{{v^2}\over 7}
 \Big].$$
 For the real case $\delta\neq 0$ we consider in any case the configurations at
 small $\delta$, but with the condition $\delta\gamma$ not small, as in [2].
 After some lengthy but straightforward calculations we obtain
 $$ v\to 1 \quad  X_{\delta}  \to {{G m^2\gamma^2}\over {\pi}}\Big[{8\over 3}-
 \delta^2 \Big({7\over 3}-\ln\uq\big(\delta^2+1/\gamma^2\big)\Big)\Big].$$
 The difference between $X$ and $X_o$ expresses, as discussed in [2], the lack 
 of compensation between the real corrections and the virtual corrections and 
 gives the rate of decay of the interference term $\cal I$ with time 
 \footnote*{The conditions that have been assumed for $\delta$ and $\gamma$ make
  the logarithm negative and so $\nu>0$ holds.}:
 $${{{\cal I}(t_1)}\over {{\cal I}(t_2)}}=\bigg[ {{t_1}\over {t_2}}\bigg]^{-\nu}
  \quad {\rm with}\quad \nu={{G m^2\gamma^2}\over {\pi}}
  \delta^2 \Big({7\over 3}-\ln\uq\big(\delta^2+1/\gamma^2\big)\Big)\Big].
  \eqno (27)$$
 Since we are dealing with gravitational effects we investigate the
 possible effect for large (macroscopic) bodies, in this case the speeds are
 certainly small, we keep simply the term proportional to $v^4$. The resulting
 expression is:
 
 $$ v\to 0 \quad  X_{\delta}  \to {{G m^2}\over {\pi}}v^4 {2\over{15}}
 (8-7\sin^2\delta )$$
 And the comparison with the same expression at $\delta=0$ gives for the
 suppression of the interference term the behaviour:
 $${{{\cal I}(t_1)}\over {{\cal I}(t_2)}}=\bigg[ {{t_1}\over {t_2}}\bigg]^{-\nu}
  \quad {\rm with}\quad
  \nu= {{G m^2}\over {\pi}}v^4 {{14}\over{15}}\sin^2\delta\;. \eqno(28)$$

\vskip 1pc
{\bf IV. Some conclusions}
\vskip 1pc
   
    The results presented here are intended to be an extension of the analogous
    investigation concerning the photon emission [2]. The low-energy 
    gravitational effects are quite negligible
\footnote*{needless to say: the high-energy gravitational effects are unknown}
    for elementary particles, anyhow the explicit computation shows a time
    decay of the interference term which is exponential in $t$, the factor that
    multiplies the time is different in the two cases: its energy
    dependence is much stronger, the angular
    dependence shows some minor differences, if fact both in gravity and in QED
    the angular dependence is logarithmic for small angles.
    It is however quite
    plausible that the same mechanism is at work for macroscopic bodies where it
    could became comparable and ever larger than the electromagnetic effects. 
    The actual expressions intended to refer to macroscopic bodies, eq.(28) may
    be questioned on the basis that the starting point was a Bloch-Nordsieck
    model designed for high energy collision, looking in detail to the procedure
    that has been used one sees that the essential point of the approximation
    was to neglect the change of velocity of the emitting body in the process of
    radiation, now it is clear that for a long wave-length emission 
    having a massive body is enough to make the variation of the velocity
    negligible also for small velocities.  
    In practice interactions with the rest of the world will be, for a 
    macroscopic body, certainly more relevant than gravitational radiation, but
    this kind of radiative process last should be quite universal and for this 
    fact it has some interest of principle.
    Another point to be stressed is the possibility that this kind of dynamics 
    acts also on photons.\par
    As it has discussed in [2] this particular decoherence effect is evidently 
    due to the presence of massless particles, and so to the possibility of 
    producing in the collision process of an indefinite number of them, so every
    exclusive channel gets, at the end, probability zero; in an S-matrix
    treatment the sum over
    different final states is not simply a practical procedure, but also a
    necessity in principle. As the radiated particles become softer and softer 
    the time needed to radiate them becomes longer and longer, for every finite 
    time no
    true i.r. divergence is present [9], the zero value for of the probability
    for every exclusive channel in anyhow attained with continuity in time.\par
    There is a more complicated effect of decoherence, which affects in case of
    superposition of two (or more) states of motion the rescattering phase, see
    eq.(7d), but the possibility of observing these effects appear even more 
    remote and also less relevant in principle, so it has not been computed.\par
   It can also be noticed that the result does not comes purely from kinematics,
   the particular form of the coupling is relevant, if e.g. one would consider a
    chiral coupling, the insertion of a massless pseudoscalar particle on an 
    external line would give
    $${1\over {\hat p-\hat k-m}}\gamma_5 u(p)={{\hat k}\over {2p\cdot k-k^2}}
    \gamma_5 u(p)$$
    and there would be no divergence in the limit $k\to 0$.\par
    Tentatively one could argue that the emission both of photons and of 
    gravitons arises from a dynamics giving a privileged role to the momentum 
    variables and therefore the persistence of states which are superposition 
    of two or more sharply different momenta is not tolerated.
    We know also that this kind of emission is related to the presence of 
    long-range
    forces [3], it is not obvious whether this has to do with the decoherence
    process, but it is certainly the reason why both the electromagnetic and the
    gravitational interaction sum up starting from microscopic size to
    macroscopic size and thus making a qualitative continuity in the
    dynamics of small and large objects.\par
    The effects of the infrared radiations are clearly due to a dynamics which
    is quite standard; this can be said beyond any doubt for the electromagnetic
    infrared emission, which is well confirmed experimentally, but also for the
    gravitational radiation, for which a direct experimental verification is not
    available, the common lore is to take for granted its existence. So these 
    effects should be there even if the quantum mechanics should undergo some 
    important modifications which, in very different forms have been postulated
    also in the aim of introducing a decoherence while keeping the relevant
    features of quantum physics. This is not the place to discuss the very large
    literature on this field, as an example one of the most carefully
    elaborated proposal is mentioned [10]. There a modification in the in the 
    evolution of every quantum system is postulated, by introducing a
    non-Hamiltonian term, while the Hamiltonian part is not modified; in the
    present treatment the
    evolution remains Hamiltonian, but in the
    limit $t\to +\infty$ the evolution operator is no longer unitary, while
    remaining isometric.
   \vskip 1pc

  {\bf Acknowledgments}
  \vskip 1pc
  I thank my colleague F.Legovini for his help. \par
  This work has been partially supported by the Italian Ministry: Ministero
  dell'Istruzione, Universit\`a e Ricerca, by means of the {\it Fondi per la
  Ricerca scientifica - Universit\`a di Trieste }.

\vfill
\eject

{\bf References}
\vskip 1pc
\item {1.} H.P.Breuer and F.Petruccione Phys. Rev. A63 032102 (2001)
\item {2.} G. Calucci Phys. Rev. A67 042702 (2003)
\item {3.} S.Weinberg Phys. Rev.140, B516 (1965)
\item {4.} F.Bloch and A.Nordsieck  Phys. Rev.52, 54 (1937)
\item {5.} J.M.Jauch and F.R\"ohrlich - {\it The theory of photons and 
           electrons,} Ch.16   Springer Verlag 1980
\item {6.} R. Aldrovandi, G.E.A. Matsas, S.F. Novaes and D. Spehler            
           Phys. Rev. D50, 2645 (1994)
\item {7.} L. Harris and L.M. Brown Phys. Rev. 105,1656  (1957)
\item {8.} D.R.Yennie and H.Suura Phys. Rev. 105,1378 (1957)  
\item {9.}  O.Steinmann  Nucl. Phys. B361, 173 (1991) 
\item {  }  O.Steinmann   Ann. Inst. H. Poincar\'e 63, 399 (1995)
\item {10.}G.C.Ghirardi, A.Rimini, T.Weber Phys.Rev.D34, 470 (1986)
\vskip 2pc

  {\centerline {\sf Figure captions}}
   \vskip 1pc
  1.) The basic $2\to 2 $ scattering graph, for particles $p,p'$ going into
  particles $q,q'$.
  \par
  2.) The four graphs for $2\to 2$ scattering with emission of a soft particle
     with very low four-momentum $k$.
  \par
  3.) Three of the six graphs for $2\to 2$ scattering with 
     virtual correction in $s,t,u$ channels, the virtual particle bears the
     four-momentum $l$.
  \par
  4.) One of the four $2\to 2 $ scattering graph with self-energy correction.

\vfill
\eject
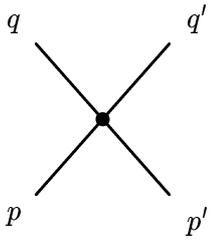
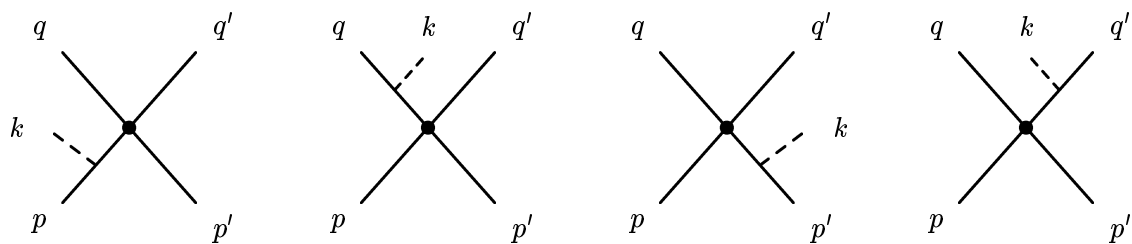
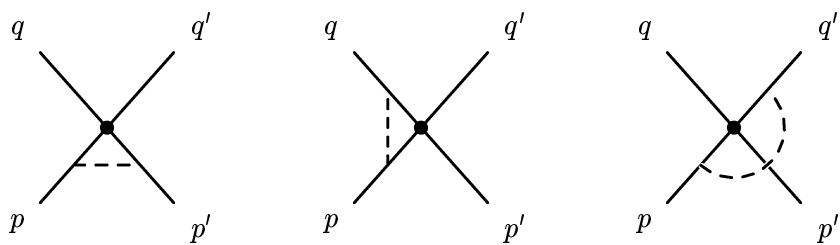
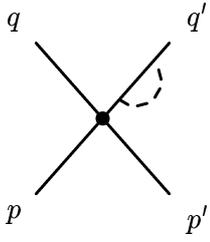
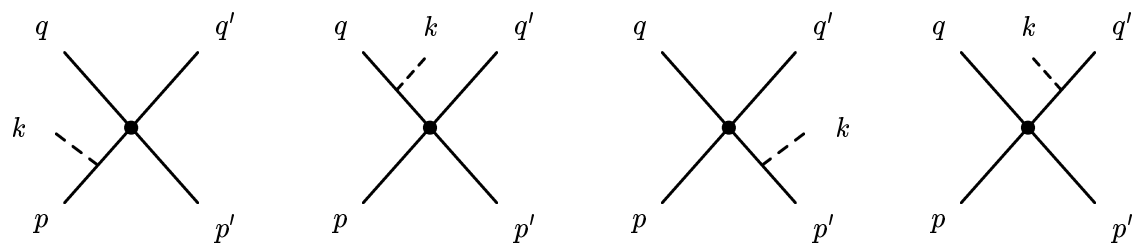
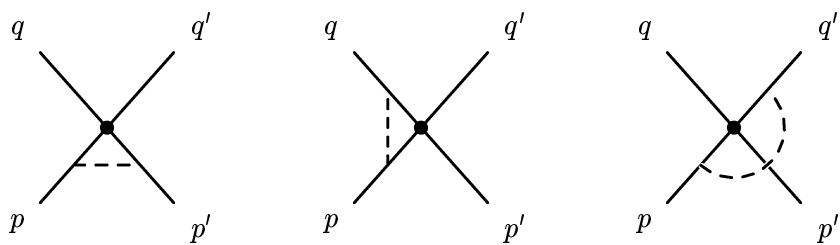
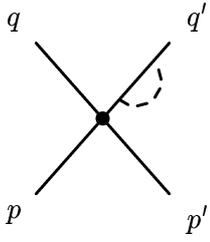
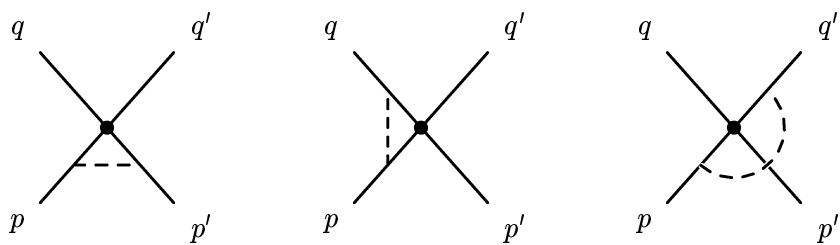
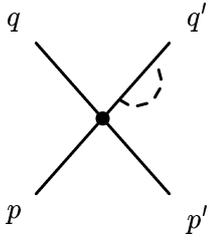
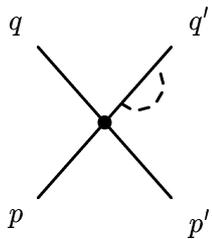
\end